\documentstyle[twoside,fleqn,espcrc2,epsf]{article}

\newcommand{\AmS}{{\protect\the\textfont2
  A\kern-.1667em\lower.5ex\hbox{M}\kern-.125emS}}
\def\lsim{\raise0.3ex\hbox{$<$\kern-0.75em\raise-1.1ex\hbox{$\sim$}}}
\def\gsim{\raise0.3ex\hbox{$>$\kern-0.75em\raise-1.1ex\hbox{$\sim$}}}

\title{
Quenched charmonium spectrum on anisotropic lattices 
\thanks{Talk presented by M.~Okamoto
}
}

\author{CP-PACS Collaboration : 
  A.~Ali~Khan\rlap,\address{Center for Computational Physics,
    University of Tsukuba, Tsukuba, Ibaraki 305-8577, Japan}
  S.~Aoki\rlap,\address{Institute of Physics,
    University of Tsukuba, Tsukuba, Ibaraki 305-8571, Japan}
  R.~Burkhalter\rlap,$^{\rm a,b}$
  S.~Ejiri\rlap,$^{\rm a}$
  M.~Fukugita\rlap,\address{Institute for Cosmic Ray Research,
    University of Tokyo, Kashiwa 277-8582, Japan}
  S.~Hashimoto\rlap,\address{High Energy Accelerator Research Organization
    (KEK), Tsukuba, Ibaraki 305-0801, Japan}
  N.~Ishizuka\rlap,$^{\rm a,b}$
  Y.~Iwasaki\rlap,$^{\rm a,b}$
  K.~Kanaya\rlap,$^{\rm b}$
  T.~Kaneko\rlap,$^{\rm d}$
  Y.~Kuramashi\rlap,$^{\rm d}$
  K.-I.~Nagai\rlap,$^{\rm a}$
  M.~Okamoto\rlap,$^{\rm b}$
  M.~Okawa\rlap,$^{\rm d}$
  H.P.~Shanahan\rlap,$^{\rm a}$\thanks{address after 15 Sept., 2000:
        Department of Biochemistry and Molecular 
        Biology, University College London, London, England, UK}
  Y.~Taniguchi\rlap,$^{\rm b}$
  A.~Ukawa$^{\rm a,b}$ and
  T.~Yoshi\'e$^{\rm a,b}$
  }

\begin{document}

\begin{abstract}
We present the results of quenched charmonium spectrum for S- and 
P-states,
obtained by a relativistic heavy quark method on anisotropic lattices.
Simulations are carried out using the standard plaquette gauge action 
and a meanfield-improved clover quark action 
at $a_t^{-1} = 3$--6 GeV with the renormalized anisotropy fixed to 
$\xi \equiv a_s/a_t =3$.
%The quark hopping parameters are tuned non-perturbatively to obtain 
%$\xi =3$ in the speed of light for a pseudoscalar dispersion relation. 
We study the scaling of our fine and hyperfine mass splittings, 
and compare with previous results.
\end{abstract}

\maketitle
\setcounter{footnote}{0}

\section{Introduction}

Conventional formalism of lattice QCD fails to describe heavy quarks
because $am_Q$ \gsim $O(1)$ for charm and bottom quarks on current 
lattices with $a^{-1} \sim 1$--3 GeV.
%
%Two widely used methods to study heavy quarks on the lattice are 
%the NRQCD method\cite{NRQCD} and the relativistic method 
%(the Fermilab method)\cite{FNAL}. 
%
%The former is based on an expansion in $1/am_Q$ and, therefore, 
%does not allow a continuum extrapolation.
%%
%The latter 
%
In heavy hadrons, however, typical scale for the spatial momenta 
are comparable with that of light hadrons. 
A momentum with $O(m_Q)$ appears only in the time direction. 
Therefore, as proposed by Klassen, it is natural to adopt an 
anisotropic lattice with a small temporal lattice spacing with 
$a_tm_Q \ll 1$ for a lattice QCD description of heavy 
hadrons\cite{Klassen98,Klassen}. 

In this report, we present results of our on-going calculation on 
the quenched charmonium spectrum using the anisotropic method, 
which attempts to test the feasibility of this approach.

\begin{table} [t]
\setlength{\tabcolsep}{0.25pc}
\begin{center}
\caption{Simulation parameters. Lattice spacing is fixed by the Sommer 
scale $r_0=0.5$~fm.}
  \begin{tabular}{cccccccc}
\hline
$\beta$ &$\xi$ &$\xi_0$ &$c_s$ &$c_t$ & $a_s^{r_0}$[fm]
&$L^3 \times T$  & $L a_s $[fm]   \\ \hline
5.7 & 3 & 2.35 & 1.97 & 2.51  &0.202 &$8^3  \times 48$ & 1.61\\
5.9 & 3 & 2.41 & 1.84 & 2.45  &0.139 &$12^3 \times 72$ & 1.67\\
6.1 & 3 & 2.46 & 1.76 & 2.42  &0.100 &$16^3 \times 96$ & 1.60\\ 
\hline
\end{tabular}
%\end{center}
%\vspace{-.7cm}
%\label{tab:simparam}
%\end{table}

\vspace{5mm}

%\begin{table} [t]
%\setlength{\tabcolsep}{0.32pc}
%\begin{center}
%\caption{Simulation parameters (2).}
\begin{tabular}{cccccc}
\hline
$\beta$  & $a_tm_{q0}$ & $\zeta$   & iter/conf
& \#conf &$c(0)$  \\ \hline
5.7   & 0.320  & 2.88  &100 &1000 & 1.009(6)  \\
5.7   & 0.253  & 2.85  &100 &1000 & 1.009(7)  \\
5.9  & 0.144  & 2.99  &100 &1000 & 0.991(8)  \\
5.9  & 0.090  & 2.93  &100 &1000& 0.991(8)  \\
%6.1  & 0.024  & 3.03  &200 & 600& 0.982(10)  \\
6.1  & 0.056  & 3.10  &200 & 600& 0.980(10)  \\
6.1  & 0.024  & 3.03  &200 & 600& 0.982(10)  \\ 
\hline
\end{tabular}
\end{center}
%\vspace{-.7cm}
\label{tab:simparam2}
\end{table}

%\section{Formalism}
\section{Method and simulation parameters}

%\subsection{Action}

The gauge action we use is given by 
\begin{equation}
S_g = \beta \sum (1/\xi_0 P_{ss'} + \xi_0 P_{st}),
\end{equation}
where $\xi_0$ is the bare anisotropy. 
The renormalized anisotropy, defined by $\xi \equiv a_s/a_t$, 
can be determined by Wilson loops. 
In this study, we use the result of \cite{calib} for 
$\xi(\xi_0,\beta)$. 

For the quark action, we employ an $O(a)$-improved 
Wilson-type quark action on an anisotropic 
lattice\cite{Klassen98,Klassen,FNAL}: 
\begin{eqnarray}
S_f &=&  \sum_{x} \{ \bar{\psi}_x \psi_x 
 \nonumber\\
&&- K_t [ \bar{\psi}_x (1- \gamma_0) U_{0,x} \psi_{x+\hat{0}} 
 \nonumber\\
&&+ \bar{\psi}_{x+\hat{0}} (1+\gamma_0) U^{\dagger}_{0,x} \psi_{x}] 
 \nonumber\\
&&- K_s \sum_i [ \bar{\psi}_x (1- \gamma_i) U_{i,x} \psi_{x+\hat{i}} 
 \nonumber\\
&&+ \bar{\psi}_{x+\hat{i}} (1+\gamma_i) U^{\dagger}_{i,x} \psi_{x}] 
\}
 \nonumber\\
&&+ i K_s c_s \sum_{x, i < j} \bar{\psi}_x \sigma_{ij} F_{ij}(x) \psi_x
 \nonumber\\
&&+ i K_s c_t \sum_{x, i} \bar{\psi}_x \sigma_{0i} F_{0i}(x) \psi_x,
\end{eqnarray}
where $K_{s,t}$ and $c_{s,t}$ are the spatial and temporal hopping
parameters and clover coefficients. The bare quark mass is given by 
\begin{equation}
a_t m_{q0} = 1/2K_t - 3/\zeta -1,
\end{equation}
where $\zeta \equiv K_t/K_s$.
We adopt the meanfield-improved clover coefficients: 
\begin{equation}
c_s = \frac{1}{\langle U_s \rangle^3}, ~~~~
c_t = \frac{1+\xi}{2} \frac{1}{\langle U_s \rangle \langle U_t
\rangle^2}_,  \label{ct}
\end{equation}
where 
$\langle U_s \rangle = \langle P_{ss'} \rangle^{1/4}$
with $P_{ss'}$ the spatial plaquette
and $\langle U_t \rangle =1$.

\begin{figure}[tb]
\vspace{-5mm}
\begin{center}
\leavevmode
\epsfxsize=7.5cm
\epsfbox{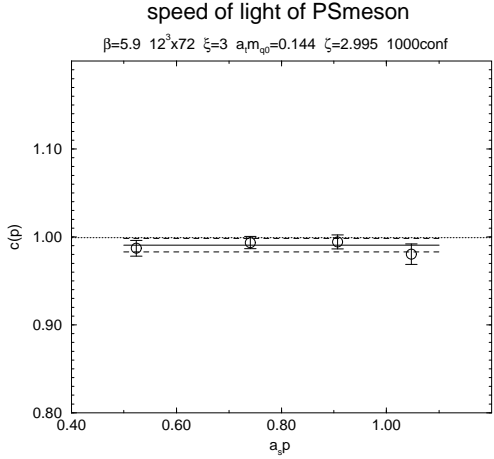}
\end{center}
\vspace{-17mm}
    \caption{The speed of light and effective speed of light 
    for $\eta_c$ at tuned $\zeta$ and $\beta=5.9$.}
    \label{cspeed}
\vspace{-15pt}
\end{figure}

%\subsection{Simulation parameters}

Our simulation parameters are summarized in Table~\ref{tab:simparam2}.
%and \ref{tab:simparam}. 
%
We study lattices with a fixed renormalized anisotropy $\xi=3$ 
and approximately the same spatial lattice size of about 1.6 fm. 
In order to study the charm quark, we simulate three lattices with 
the lattice cutoff in the range $a_t^{-1} = 3$--6 GeV. 

The parameter $\zeta$ has to be tuned such that 
the relativistic errors are eliminated.
%anisotropy in the quark sector coincides with that in the gluon sector. 
%
For this purpose preparatory simulations are 
performed at $\zeta=2.8$, 3.0 and 3.2.
We measure the pseudoscalar meson mass at four lowest on- and off-axis 
momenta and perform a fit of form, 
\begin{equation}
E(p)^2 = E(0)^2 + c^2(p)p^2
\label{EPrel}
\end{equation}
to extract the ``speed of light'', $c(p)$, using three
or four lowest momenta.  The tuning condition for 
$\zeta$ is $c(p=0) = 1$ which requires the relativistic dispersion
relation to be restored for small momenta. 
(For massless free fermions, $c(p=0) = 1$ is satisfied at 
$\zeta=\xi$.)
Results for $\zeta$ are summarized in Table~\ref{tab:simparam2}.

Figure~\ref{cspeed} shows a typical result for the effective speed of light
$
c_{\rm eff}(p) = \sqrt{E(p)^2 - E(0)^2}/p, 
$
obtained by a simulation made with a tuned value of $\zeta$. 
The wide plateau indicates that the linearity of $E(p)^2$ in $p^2$ is
well satisfied. 
The horizontal solid line is the result of $c(0)$ from a global fit 
according to (\ref{EPrel}), where the dashed lines indicate its error. 
We see that $c(0)$ is consistent with 1.
As summarized in Table~\ref{tab:simparam2}, 
the condition $c(0)=1$ is confirmed to be satisfied within 1--2\% 
with our values of $\zeta$.

\begin{table} [t]
\setlength{\tabcolsep}{0.2pc}
\begin{center}
\caption{S- and P-state operators.}
  \begin{tabular}{ccccl}
\hline
state & $J^{PC}$ & name & local & non-local   \\ \hline
$^1S_0$ & $0^{-+}$ & $\eta_c$ & $\bar{\psi} \gamma_5 \psi$ &  \\
$^3S_1$ & $1^{--}$ & $J/\psi$ & $\bar{\psi} \gamma_i \psi$ &\\
$^1P_1$ & $1^{+-}$ & $h_c$ & $\bar{\psi} \sigma_{ij} \psi$ &
 $\bar{\psi} \gamma_5 \Delta_i \psi$   \\
$^3P_0$ & $0^{++}$ & $\chi_{c0}$ & $\bar{\psi}  \psi$ &  
$\bar{\psi} \sum_i \gamma_i \Delta_i \psi$ \\
$^3P_1$ & $1^{++}$ & $\chi_{c1}$ &$\bar{\psi}
\gamma_i\gamma_5 \psi$   & $\bar{\psi} \{ \gamma_i \Delta_j - \gamma_j \Delta_i
\} \psi$  \\
$^3P_2$ & $2^{++}$ & $\chi_{c2}$ &  &   $\bar{\psi} \{ \gamma_i
\Delta_i - \gamma_j \Delta_j \} \psi$ (E)\\
 & & & & $\bar{\psi} \{ \gamma_i \Delta_j + \gamma_j
\Delta_i \} \psi$ (T)\\ \hline
\end{tabular}
\end{center}
\label{tab:SPope}
\end{table}

\subsection{Measurements}

At each $\beta$, hadronic measurements are made on configurations 
separated by 100--200 iterations, 
at two values of $m_{q0}$. 
See Table~\ref{tab:simparam2}.
These two quark masses are chosen such that the charm quark point 
can be interpolated.

We measure all S- and P-state mesons using both local and non-local
operators compiled in Table~\ref{tab:SPope}.
For the non-local operators we adopt an exponentially smeared
derivative sources: 
\begin{equation}
f_i({\bf x}) = A_s e^{-B_s \mid {\bf x}-\hat{i}\mid} - 
             A_s e^{-B_s \mid {\bf x}+\hat{i}\mid} 
~~~ (i=1,2,3), 
\end{equation}
where $A_s$ and $B_s$ are the smearing parameters.  
We extract meson masses by a single cosh fit. 
Errors are determined by a jack-knife method.

\begin{figure}[t]
\vspace{-5mm}
\begin{center}
\leavevmode
\epsfxsize=7.5cm
\epsfbox{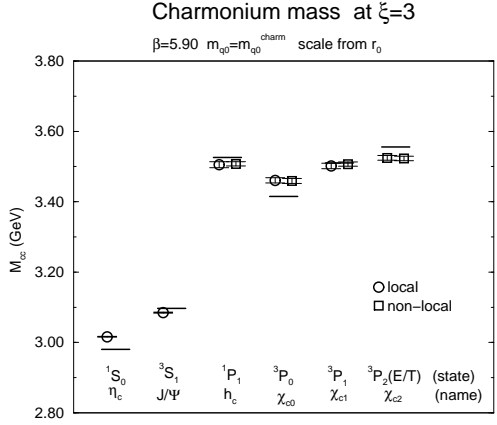}
\end{center}
\vspace{-18mm}
    \caption{The charmonium mass spectrum at $\xi =3$ and
    $\beta=5.9$. The scale is fixed through the Sommer scale 
    $r_0$.}
    \label{massplot}
\vspace{-15pt}
\end{figure}

\begin{figure}[t]
\vspace{-5mm}
\begin{center}
\leavevmode
\epsfxsize=7.5cm
\epsfbox{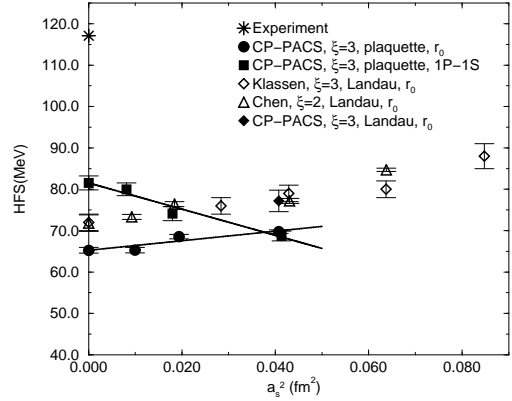}
\end{center}
\vspace{-18mm}
    \caption{The scaling behavior of hyperfine-splitting
    $\Delta M_{J/\psi - \eta_c}$. Filled circle and square
    are our results with scale from the Sommer scale $r_0$ and
    the $1P - 1\bar{S}$ splitting respectively. Open diamond and 
    triangle are results of Ref.\cite{Klassen} and
    \cite{Ping}.
%    with the Landau mean link meanfield estimates for clover
%    coefficients. 
%    These captions also apply to
%    the figures that follow.
    }
    \label{HFS}
\vspace{-8pt}
\end{figure}

\begin{table}[bt]
\vspace{-7mm}
\begin{center}
\caption{Preliminary continuum estimates of mass splittings
and the experimental values.}
  \begin{tabular}{cccc}\hline
 & $a_s \rightarrow 0$   & $a_s \rightarrow 0$  & Exp. \\ \hline
scale & $r_0$ & $^1P_1 - \bar{S}$   &\\  \hline
$\Delta M_{J/\psi - \eta_c}$[MeV] & 65(1) & 82(2)  & 117\\
$\Delta M_{\chi_{c1}- \chi_{c0}}$[MeV] & 55(5) & 63(5) & 95\\
$\Delta M_{\chi_{c2}- \chi_{c1}}$[MeV] & 19(7) & 21(7)    & 46\\
$\frac{\Delta M_{\chi_{c2}- \chi_{c1}}}{\Delta M_{\chi_{c1}-
\chi_{c0}}}$ & 0.33(15) & 0.31(14) &  0.48\\ \hline
\end{tabular}
\end{center}
\vspace{-1.0cm}
\label{tab:rst}
\end{table}

\begin{figure}[tb]
\vspace{-10mm}
\begin{center}
\leavevmode
\epsfxsize=7.5cm
\epsfbox{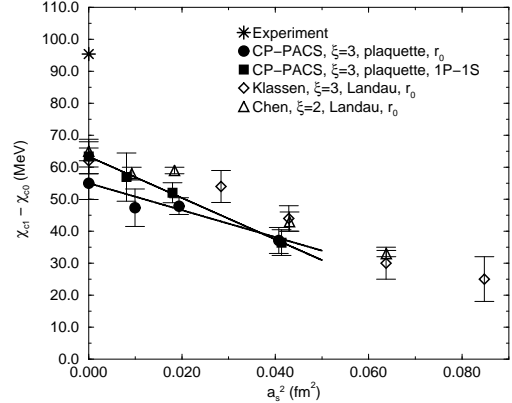}
\end{center}
\vspace{-18mm}
    \caption{The scaling behavior of $\Delta M_{\chi_{c1}- \chi_{c0}}$. }
    \label{fine1}
\vspace{-15pt}
\end{figure}

\begin{figure}[thb]
\vspace{-5mm}
\begin{center}
\leavevmode
\epsfxsize=7.5cm
\epsfbox{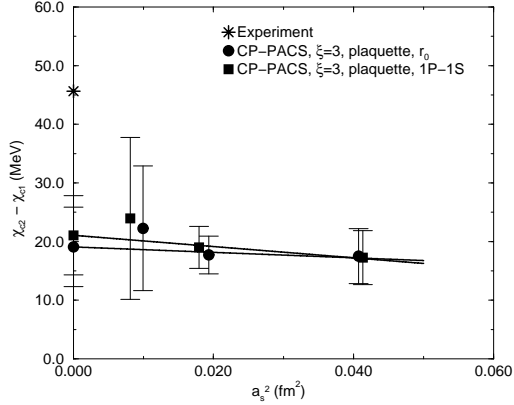}
\end{center}
\vspace{-18mm}
    \caption{$\Delta M_{\chi_{c2}- \chi_{c1}}$.}
    \label{fine2}
\vspace{-15pt}
\end{figure}

\begin{figure}[tb]
\vspace{-5mm}
\begin{center}
\leavevmode
\epsfxsize=7.5cm
\epsfbox{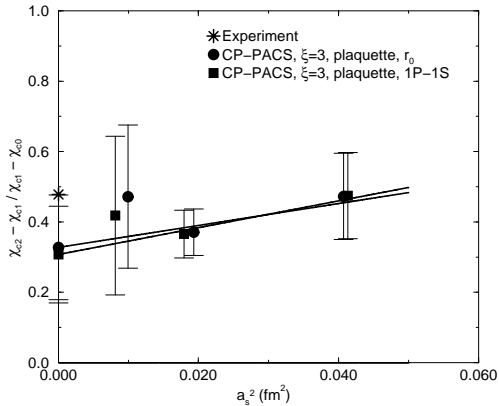}
\end{center}
\vspace{-18mm}
    \caption{The ratio $\Delta
    M_{\chi_{c2} - \chi_{c1}}/\Delta M_{\chi_{c1}-\chi_{c0}}$.}
    \label{ratio}
\vspace{-15pt}
\end{figure}

\section{Charmonium spectrum}

We fix the scale using either
%string tension $\sigma$ 
the Sommer scale $r_0=0.5$~fm or the $1^1P_1 - 1\bar{S}$ splitting.
The charm quark point is then defined by setting the spin averaged 
1S meson mass ($1\bar{S}$). 
Figure~\ref{massplot} is a typical result for the charmonium mass spectrum 
obtained at $\beta=5.9$. 
%In this figure, the scale is fixed by the Sommer scale $r_0$. 
%
In the following, we study the lattice spacing dependence of 
the spectrum. 
With the $O(a)$ improved action, the leading scaling violation is
expected to be proportional to $a^2$. 

\subsection{Hyperfine splitting}

Figure~\ref{HFS} is our result for the S-state hyperfine-splitting 
$J/\psi - \eta_c$ as a function of the lattice spacing. 
Results using the scales from $r_0$ and $1^1P_1 - 1\bar{S}$ are 
presented. 
We find that, although our results for the hyperfine-splitting can 
be approximately fitted by a linear function of $a_s^2$, 
the two results from different scales lead to different values even 
in the continuum limit. 
This discrepancy should be regarded as a quenching artifact.

Plotted together are the previous lattice results by 
Klassen\cite{Klassen} and Chen\cite{Ping} at $\xi=2$ and 3, 
using the same lattice action.
A difference between our simulation and those in 
\cite{Klassen} and \cite{Ping} is the choice of the tadpole factor
for the clover coefficients; 
we use the fourth root of the plaquette expectation value,
while the mean link in the Landau gauge is used in \cite{Klassen} and 
\cite{Ping}. 
To check the consistency, we also performed a simulation at 
$\beta =5.7$ using the Landau gauge mean link for the meanfield.
As shown in Fig.~\ref{HFS} by a filled diamond, our result is consistent 
with those by Klassen. 

We find that our result from the scale $r_0$ is approximately parallel 
to the other two results, and lead to a continuum limit 
which is about 10\% lower than that of the others. 
The origin of this discrepancy is not clear to us at present.

\subsection{Fine structure}

Results for the P-state fine structure $\chi_{c1}- \chi_{c0}$ are 
shown in Fig.~\ref{fine1}. 
Although the scaling behavior of our results is similar to those 
of \cite{Klassen} and \cite{Ping}, the continuum limit deviates 
by about $2\sigma$ for the data using the $r_0$ scale. 

We also show the results for 
the $\chi_{c2}- \chi_{c1}$ splitting and the ratio $\chi_{c2}-
\chi_{c1}/\chi_{c1}-\chi_{c0}$ in Figs.~\ref{fine2} and \ref{ratio}. 
No data from other groups are available for these quantities.

\section{Conclusions}

We have presented our results for the charmonium mass 
spectrum obtained on an anisotropic lattice with $\xi=3$. 
Our results for fine and hyperfine mass splittings are shown in 
Figs.\ref{HFS}-\ref{ratio}. 
Our preliminary values from a continuum extrapolation linear in $a^2$
are summarized in Table~3.
%\ref{tab:rst}. 
We find that these results are much smaller than experimental values.

We also find that the results are quite sensitive to the choice of the 
clover coefficient. 
A naive continuum extrapolation, linear in $a^2$, does not resolve 
the discrepancy between different choices of the clover coefficient.
We are now extending the simulation to a finer lattice, to examine 
the reliability of continuum extrapolations.

\vspace*{3mm}
%\section*{Acknowledgment}

This work is supported in part by Grants-in-Aid
of the Ministry of Education ( Nos. 
10640246, 10640248, 10740107, 11640250, 11640294, 11740162,
12014202, 12304011, 12640253, 12740133).
AAK is supported by JSPS Research for the Future Program
(No. JSPS-RFTF 97P01102).
SE, TK, KN, M. Okamoto and HPS 
are JSPS Research Fellows.

\end{document}